# AI/ML for 5G and Beyond Cybersecurity

Sandeep Pirbhulal, Habtamu Abie, Martin Jullum, Didrik Nielsen, and Anders Løland
Norwegian Computing Center, P.O.Box 114, Blindern, NO-0314 Oslo, Norway
{sandeep, abie, jullum, anderslo}@nr.no

**Abstract:** The advancements in communication technology (5G and beyond) and global connectivity Internet of Things (IoT) also come with new security problems that will need to be addressed in the next few years. The threats and vulnerabilities introduced by AI/ML based 5G and beyond IoT systems need to be investigated to avoid the amplification of attack vectors on AI/ML. AI/ML techniques are playing a vital role in numerous applications of cybersecurity. Despite the ongoing success, there are significant challenges in ensuring the trustworthiness of AI/ML systems. However, further research is needed to define what is considered an AI/ML threat and how it differs from threats to traditional systems, as currently there is no common understanding of what constitutes an attack on AI/ML based systems, nor how it might be created, hosted and propagated [ETSI, 2020]. Therefore, there is a need for studying the AI/ML approach to ensure safe and secure development, deployment, and operation of AI/ML based 5G and beyond IoT systems. For 5G and beyond, it is essential to continuously monitor and analyze any changing environment in real-time to identify and reduce intentional and unintentional risks. In this study, we will review the role of the AI/ML technique for 5G and beyond security. Furthermore, we will provide our perspective for predicting and mitigating 5G and beyond security using AI/ML techniques.

**Keywords:** 5G, 6G, Cybersecurity, AI, ML

# 1 Introduction

This report aims to provide a comprehensive overview of the challenges that how AI/ML techniques can be used for 5G and beyond cybersecurity based on studies published until the end of 2022. Also, this report includes challenges that AI/ML techniques face in protecting cyberspace against attacks in 5G and beyond IoT applications.

The research questions (RQs) are**:**
RQ1. What are the cybersecurity challenges, attack scenarios, risk management methodologies for the 5G and beyond enabled IoT applications?
RQ2. What are the new challenges to 5G and beyond to support cybersecurity for IoT applications?
RQ3. How can AI/ML technologies be used in cybersecurity to construct smart models for defending systems from attacks in 5G and beyond enabled IoT applications?

To answer these RQs, this report will have the following main objectives:
O1. Focus on the fifth and sixth generation and shed light on various cybersecurity aspects.
O2. Study the challenges to be overcome in 5G and beyond to better support the IoT applications, including network architectures, network resource allocation schemes, and advanced techniques, etc.
O3. Survey on deep learning and AI techniques as promising approaches to unleash the full potential of Cybersecurity in 5G and beyond networks, IoT and cyber-physical systems. It includes:

- AI/ML enhancing cybersecurity beyond 5G
- Securing AI/ML algorithms
- Threats and vulnerabilities of AI/ML algorithms
- Mitigating measures
- Classes of data being used (tabular, time series, text, …)
- The occurrence of supervised/semi-supervised/unsupervised learning

The rest of this report is organized as follows: Section 2 includes the 5G cybersecurity challenges, potential 5G vulnerabilities and risks, and 5G threat landscape. Section 3 elaborates on the understanding of 5G and beyond, the evolution of the communication network security landscape from 1G to 6G, potential vulnerabilities and risks and security challenges for 5G and beyond. In Section 4, the AI/ML cybersecurity in 5G and beyond networks is discussed. The concluding remarks and future research directions are discussed in Section 5.

# 2  5G Cybersecurity

5G networks aim at delivering value-added services with evolved performance such as low-latency communications, high data rates, increased reliability and capacity to sustain more interconnected devices. 5G is more evolved than 4G mobile communication, comprising protocol, speed, and network configurations. Network Function Virtualisation (NFV) and Software-Defined Networking (SDN) concepts have been introduced to fulfil the necessities for scalability, throughput, latency, and reliability. On one hand, the NFV and SDN offer flexible networks and the rapid development of services. On the other hand, they are more prone to security due to softwarization and virtualization, thus causing additional security challenges and complexities in some cases [Afaq 2021, Hussain 2021].

Recent ENISA's report [ENISA 2022] covers the increased attack space introduced using virtualized functions such as Virtual network functions (VNFs) and Virtual Machines (VMs) in 5G. They argue that the rapid development of cloud native technology (CNT) advances the telecommunication infrastructure as 5G networks require virtualisation and CNT to support interworking with existing networks, and to support both VMs and containers at the same time. Several tenants can share the dynamic NFVs and devices for different deployment scenarios with varying risks, such phenomena may expose the level and severity of security risks on several different levels. The report addresses some possible security challenges in 5G due to weak isolation among resources in different VNFs such as side channel attacks, container data theft, illegitimate access to other VNFs, etc.

Seongmin et al. [Seongmin 2021] discussed that the global 5G security market is expected to increase by 50% (USD 4 billion) in 2023 and about USD 7 billion in 2025. Since most of 5G service providers offer 5G services without solutions to security threats, robust and efficient security approaches are required for 5G [Chlosta 2019, Kim 2019].

## 2.1  5G Cybersecurity Challenges

5G needs robust security solutions since it will combine several real-time applications with communication networks. The important security challenges in 5G are investigated and pinpointed in [Ijaz 2018, Hassan 2022]. The fundamental challenges in 5G are as follows [Hassan 2022]:

- DoS attacks on end-user devices
- Flash network traffic
- User plane integrity
- Roaming security
- Denial of Service (DoS) attacks on the infrastructures
- Security of radio interfaces
- Signalling storms

However, due to the introduction of virtualization and softwarization in 5G, there can be many other security threats at NFVs. NFVs provide flexible networks and prompt service creation while introducing supplementary security challenges and complexities. It is essential to address the security issues and to focus on instilling more robust security and privacy settings in 5G NFV systems in the following areas [ENISA 2022]:

- Virtualisation or containerisation: Major security challenges are posed due to SDNs' susceptible to attacks such as forwarding device attacks, control pane threats, API vulnerabilities, counterfeit traffic flows and more, and dynamic distribution of NFV architectures.
- Orchestration and management: Security challenges such as vulnerabilities within orchestration protocols, Management and Orchestration (MANO) single point of failure, orchestration compromise and policy violations etc. can incur due to a lack of consistency on how to manage and orchestrate the network services.
- Administration and access control: 5G NFV network architecture offers openness and programmability relying on the expanded use of APIs. Network functions and sensitive parameters with inaccurate access control rules may be exposed due to improperly designed or configured API.
- New and legacy technologies: The communication between physical and virtual environments in 5G may also raise several security challenges at management and orchestration layers which need to be considered.
- Adoption of open source: Another security challenge related to the use of lower cost, commercial off-the-shelf (COTS) hardware for network functions based on NFV technologies which may impact security and performance.
- Supply chain: The 5G NFV supply chain is susceptible to risks such as malicious software and hardware, counterfeit components, poor designs, manufacturing processes, and maintenance procedures.
- Lawful Interception (LI): Another security challenge is securing and hiding LI functions from other functions in an NFV environment.

***Analysis of the current 5G Cybersecurity Challenges:***

As mentioned above, it has been argued by different researchers that the deployment of 5G will extend various networking opportunities because of capacity, speed, and latency improvements. However, these advancements also unlock the potential for additional security attacks in 5G. Various studies [Hassan 2022, ENISA 2022, Afaq 2021, Hussain 2021] have discussed several of the cybersecurity challenges in 5G. From our analysis, one of the most significant security concerns in 5G is the possibility of DDoS attacks. DDoS attacks usually target linked devices, infrastructures and applications. As the 5G network expands, DDoS threats could target more attacks in terms of frequency and extent to which attacks may spread. DDoS attackers may maliciously try to disturb the availability of 5G devices by introducing large volumes of false packets or requests to overwhelm the network. These attacks may flood the 5G network infrastructure with fraudulent traffic, thus swallowing all available bandwidth of the network. In recent studies [Kim 2022, Saha 2022] mentioned that the optimal features selection can play a vital role in DDoS attack classification using ML or DL models. Saha et al. [Saha 2022] developed ML based approach by performing various experiments on feature selection to decrease the time complexity of detecting and examining real-time DDoS attacks in the 5G core network. Thus, it will be interesting to investigate how AI can be used for the mitigation of DDoS attacks in 5G systems before they happen. Also, it is worth trying to experiment that how AI can be useful in solving new security challenges in NFV and SDN.

## 2.2  5G Threat Landscape

As stated in [Marco 2019], the traditional threats and additional new 5G network (core, access and edge) threats are included in the 5G threat landscape. The high-level categorization of threats based on ENISA threat taxonomy [ENISA 2022] are:
- Nefarious activity/abuse (NAA)
- Eavesdropping/Interception/ Hijacking (EIH)
- Physical attacks
- Unintentional or Intentional Damage of property or persons due to failures
- Unexpected disruptions of service
- Disasters such as a sudden accident or a natural catastrophe
- Legal actions of third parties (contracting or otherwise) to forbid activities or recompense for loss as per law

Along with the above-mentioned general taxonomy, the security threats can be classified layer wise such as core network, radio access, network virtualisation or generic infrastructure component as follows:
- *Core Network threats*: These threats are related to SDN, NVF, NS and MANO, and lie in the categories of NAA and EIH.
- *Access network threats*: These threats are related to radio access technology, radio access network and non-3GPP access technologies, and lie in the categories of EIH.
- *Multi-edge computing threats*: These threats are also related to NAA and EIH categories but mostly associated with components located at the edge of the network.
- *Virtualisation threats*: These threats are associated with NFVs and VNFs operations and functions.
- *Physical Infrastructure threats*: These threats affect IT infrastructures that support the network and lie in the categories of physical attacks, damage or loss of equipment, disruption of services, and disasters.
- *Generic threats*: These threats such as Denial of Service (DoS) typically affect any ICT system or network
- *SDN threats*: These threats are related to the softwarization functions that are the backbone of 5G.

Additionally, the attacks on 5G NFV can be categorized as follows (Figure 1):
- attacks from within an NFV
- attacks from outside an NFV
- attacks occurring between NFV components

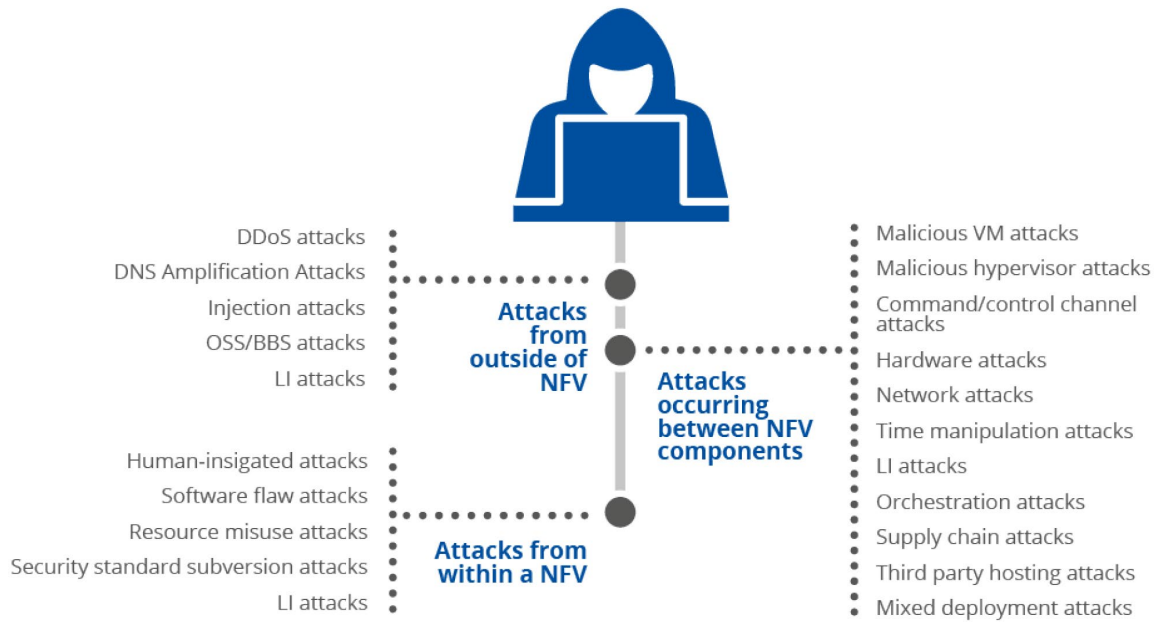
Figure 1 - Attack Taxonomy [ENISA 2022]

***Analysis of the current 5G Threat Landscape:***
5G will be the backbone of cellular networks in the coming few years; thus, service providers need to take sensible steps for 5G network protection. One major security threat is the unexpected disruption of services [ENISA 2022]; operators should consider solving the signaling challenges that may result in increased traffic volumes in 5G network deployments. The legacy security issues related to service disruption in 3G/4G (such as credential theft, location tracking, subscriber fraud, etc.) must also be addressed as service providers migrate to 5G. Since 5G will initially be using a 4G core network, service providers must address previous security concerns. Through monitoring of available signaling data crossing the border of their traditional network may assist service providers in developing protected 5G services.

As it has been argued by various authors that there will be some new threats in 5G along with existing security challenges. From the operators and equipment vendors perspective, it will be good to do research on the detection of unknown threats in 5G environments using ML or DL models.

## 2.3  5G Vulnerabilities and Risks
Cybersecurity vulnerabilities might provide opening for attackers to exploit 5G systems causing severe damages. Hassan et al. [Hassan 2022] argued that the principal vulnerabilities that may cause security challenges are:
- Pre-authentication traffic
- Jamming attacks
- Insecure inter-network protocols
- Insecure implementations in network components
- Signalling-based DoS attacks

The 5G is vulnerable to cyber threats in five possible ways [Tom 2019]:
1. In the 5G, the network has shifted from centralized and hardware-based switching to distributed and software-based routing. In the previous networks, hardware choke points were used as initial points for attack inspection. But in the 5G, SDN is used for digital routing, thus disallowing chokepoint inspection and control.
2. Due to the application of virtual functions at a higher level in 5G, all routing activities are done using IP protocols which are vulnerable and can be used for illegal activities in the network.
3. Although it is possible to lock down software vulnerabilities within the network, the network is managed by software which can be vulnerable and an attacker which can gain control of this software can also control the network.
4. In 5G, an increased bandwidth creates additional attack surface, and the small-cell and short-ranged antennas could be new targets for attackers. These cell sites use dynamic spectrum sharing capability in which each slice has varying risks. Thus, cyber protections which aren't dynamic could be vulnerable to dynamic attacks.
5. Various smart devices are attached to IoT, which are easily hackable and uniquely vulnerable. Hackers can discover insecure IoT devices into which they might plant exploitation software.

In [Vaibhav 2021], the cybersecurity risks in 5G are due to the following:
- *Data Sharing*: The critical area of risk under 5G is data compromise; it is due to the raised spread of intermediaries in the network and the software, including getting access to the data network. This poses a considerable threat to the privacy of the users.
- *Connectivity Compromise*: The connectivity compromise, also known as availability compromise, might cause vital consequences on national security.
- *Speed is a two-edged sword*: 5G provides services in very fast speed but this also means that data can be stolen very fast and data compromise can occur in high-speed.
- *Internet of Things (IoT)*: By increasing the number of users in a single network might increase the possible threats and vulnerabilities, because unauthorized users can access other users' data via shared spectrum. In addition, different devices can be connected to the same system which makes ensuring cybersecurity in each device more complex.
- *Lack of end-to-end encryption*: Multi communication channels connecting 5G with Wi-Fi have lack of end-to-end encryption which allows attackers to intercept data.
- *Short Range Towers*: High number of Short-range physical cell towers are required in the infrastructure of 5G network which becomes new deployment of physical targets and have dynamic spectrum sharing (DSS) that hackers can exploit.
- *Unaddressed past inefficacies*: Issues in previous generations of mobile technology have not been addressed in 5G standards. One example is the ability to intercept pre-authenticated messages between the user's base station and cell tower.

**Analysis of the current 5G Vulnerabilities and Risks:**
In 5G networks, there are serious security issues in accessing radio networks because of new vulnerabilities which may affect both the service providers and end users. The core network and radio access capabilities may be obtained from the end users without following proper authentication process. This can be one of the key vulnerabilities in 5G which allows an active or passive attackers to steal the identity of devices [Shaik 2021]. This vulnerability is derived from traditional networks (3G/4G). Thus, 5G networks must come up with potential solutions for this. In 5G, another important vulnerability is when the service providers are requesting the radio access capabilities from the end device before the radio resource control (RRC) protocol security setup. Consequently, end devices' information is sent to the base stations in plain-text and there may exist an attacker who may steal the sensitive information. Thus, it will be good to see how fuzzy logic or advanced data analytics can be used to analyze the new vulnerabilities in 5G networks while considering the dynamic and scalable properties of edge and IoT.

## 3   5G and Beyond Security

The evolution in terms of communication network security is shown in Figure 2. The advantages of 5G and beyond are the unique combination of high-speed connectivity, low latency, and more effectiveness and efficiency than its predecessor technologies. 6G is expected to allow high data rates at one terabyte per second and to support more mobile connections than 5G capacity which will enable applications and innovations in connectivity, cognition, sensing, and imaging with wireless networks empowered by Artificial Intelligence (AI) [Peltonen 2020, Wang 2020].  These new areas are prone to security and privacy issues such as authentication, access control, confidentiality, malicious behavior, encryption and data transmission [Wang 2020] and performance of AI/ML models and their accuracy, accountability and trustworthiness.  On one hand correct application of AI/ML can enhance privacy. On the other hand, privacy violations may occur on AI/ML attacks on training (e.g., poisoning attack) and testing phases (e.g., reverse, membership interference, adversarial attacks).

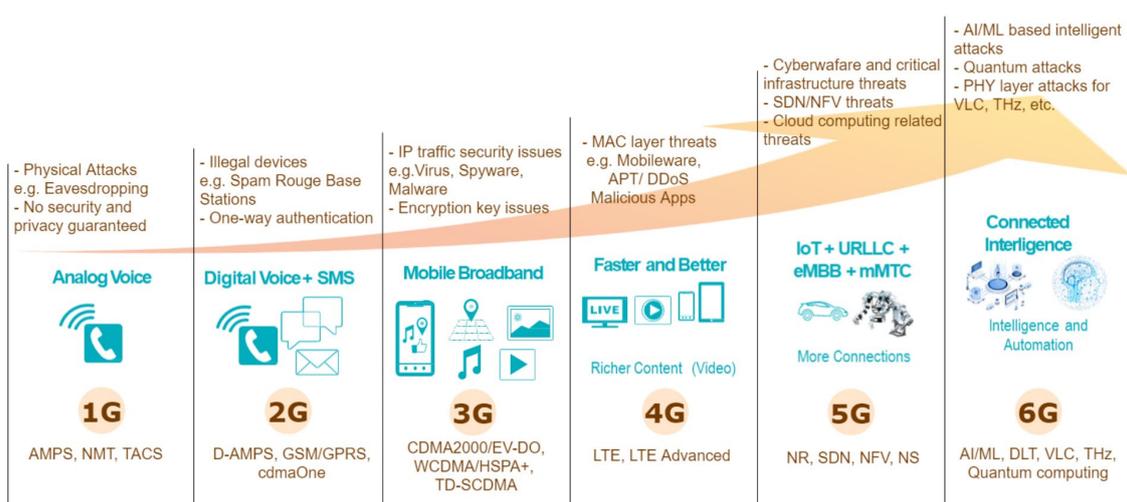

Figure 2 - Evolution of communication network security landscape [Pawani 2021]

### 3.1   Understanding 5G and Beyond Security

There exist surveys on 6G security and privacy issues from different perspectives, key aspects of 6G networks (real-time intelligent edge computing, distributed artificial intelligence, intelligent radio, and 3D intercoms) [Wang 2020], 6G wireless systems and requirements [Porambage 2021-1, Porambage 2021-2], 6G prospective technologies (physical, connection, and service layers) [Nguyen 2021], distributed ML for 5G and beyond [Nassef 2022], trust networking for beyond 5g and 6G [Kantola 2020], network architecture [Shokoor 2022], key areas beyond 5G toward 6G network [Lavanya 2022], etc. In the ensuing section we give a brief review of these surveys.

In the perspective of key aspects of 6G networks, Wang et al. [Wang 2020] survey the 6G security and privacy issues in four key aspects of 6G networks, (i) real-time intelligent edge computing, (ii) distributed artificial intelligence, (iii) intelligent radio, and (iv) 3D intercoms. The authors discussed some promising emerging technologies in each area with an overview of the security and privacy issues associated with each technology.

In the area of 6G wireless systems and requirements, Porambage et al. [Porambage 2021-1, Porambage 2021-2] survey the impact of security and privacy on 6G wireless systems and 6G requirements such as novel network architecture, applications and enabling technologies including distributed ledger technologies, physical layer security, distributed artificial intelligence (AI)/ machine learning (ML), Visible Light Communication (VLC), THz bands, and quantum communication. The authors provided their vision on the new threat landscape expected for these networks and evolving security solutions and technologies to protect 6G networks.

In the area of 6G prospective technologies, Nguyen et al. [Nguyen 2021] provide a systematic overview of security and privacy issues in the physical, connection, and service layers, as well as through lessons learned from the failures of existing security architectures and state-of-the-art defenses. They described two key lessons learned as follows. First, other than inheriting vulnerabilities from the previous generations, 6G has new threat vectors from new radio. Second, the top promising techniques to mitigate the attack magnitude and personal data breaches are physical layer protection, deep network slicing, quantum-safe communications, artificial intelligence (AI) security, platform-agnostic security, real-time adaptive security, and novel data protection mechanisms such as distributed ledgers and differential privacy.

In the area of distributed ML for 5G and beyond Nassef et al. [Nassef 2022] argue that 6G is expected to have new capability for ubiquitous communication, computation, sensing and controlling for a variety of sectors, which will increase the complexity in a more heterogeneous environment. This increased complexity makes the application of ML and distributed ML necessary. They gave graph models as examples to further improve communication and computation efficiency and scalability of the distributed ML algorithms.

In the area of trust networking for beyond 5G and 6G, Kantola [Kantola 2020] argues that trust networking is emerging as a new promising type of technology and service that could be deployed for special services over 5G or 6G networks to providing a much more predictable level of service as compared with today's service. This will help implement trustworthy use cases for safety critical mobile networks which create lots of privacy sensitive information on the physical world/ digital world boundary.

In the area of network architecture, Shokoor et al. [Shokoor 2022] in their "Overview of 5G & Beyond Security" describe 8 security landscape threats for 5G network architecture and identify possible attack entry points for each threat in the network. They analyze the 5G security model, 5G threat setting, IoT threat scenarios, and 5G network threats with a detailed review of security problems in core 5G security areas including access control, authentication, communication protection, and encryption.

In key areas beyond 5G toward 6G network, Lavanya et al. [Lavanya 2022] surveyed 6G network security and privacy by considering key areas such as intelligent radio, distributed artificial intelligence, real-time intelligent edge, and intercom 3D, and countermeasures to overcome these issues. The authors proposed a new design idea based on the deep learning method (CNN-LSTM) for implementation in a 6G network for authentication by detecting the attack in network traffic.

***Analysis of the 5G and Beyond Security:***
Most of the existing surveys on AI/ML security mainly focus on the performance of AI/ML models and their accuracy, but they often overlook the accountability and trustworthiness of the models' decisions. 6G security approaches must allow transparent and understandable security decision-making procedures pushing the methods accountable for automatic measures. Since security, privacy, and trust are closely related and will be critically important to the uptake and acceptance of 6G communication, any security solution should take a holistic approach addressing these three aspects.

## 3.2 Potential 6G cybersecurity challenges

6G is the network facilitator for a wide range of new applications to reshape the society of the 2030s and beyond. Nevertheless, these applications may face complex security challenges because of their critical nature and the requirement for a high trust level [Porambage 2021]. 6G applications may also have distinctive vulnerabilities, for example, the connected autonomous systems mainly rely on AI and the VLC technology, where malicious behaviour, encryption and data sending can be challenging [Minghao 2021].

One of the main trends in developing 6G networks is realizing network functions from a software perspective rather than hardware. Virtualization can create another level of abstraction by providing a framework for deploying network functions as software components, improving cost-effectiveness and end-to-end reliability. The security functions implemented in the software can be exploited, which requires enhanced network security measures [Giordani 2020].

An autonomous, self-aware and adaptive system is required to provide a large amount of data analysis for 6G network security by using the latest algorithms and recent AI technologies. Active security measures based on AI should continuously collect and analyse real-time data to reduce possible cyber risks and ensure network security [Tang 2019]. 6G will be pivotal for intelligent medical, industry 4.0, and other IoT applications. Along with 6G security, data and user privacy needs attention which keeps varying in the different application requirements. Since 6G will deliver ubiquitous digital services, these applications will yield massive data, thus letting more assailants impose threats on users' privacy [Pham 2020].

***Analysis of the potential 6G cybersecurity challenges:***
Various studies ([Pham 2020], [Xu 2022], [Hireche 2022]) have discussed that 5G networks have started using AI as their backbone in the physical layer, data layer, network layer (some applications). It has been also argued by some researchers ([Ioannou 2022], [Yang 2022]) that 5G and beyond (6G) will consider highly distributed AI, moving the intelligence from the central cloud to edge computing resources. It has been analyzed from the literature that in the 6G there will be narrowly controlling of data transmission between the source and destination nodes, reducing the probability of eavesdropping via the communication channel. However, there is a vast potential for man-in-the-middle attacks directed at 6G frequencies which may be the critical security concern. The attackers may introduce the metasurface to understand the communications at 6G frequencies by positioning it straight in the line of sight between end users. Thus, one open domain to do research in 6G can be man-in-the-middle attack intrusion detection using AI/ML.

### 3.3 5G and Beyond Threat landscape

In this section, 5G and beyond threat landscapes is discussed. 5G and beyond security threats are divided into three categories i) pre-6G security issues explain the security threats inherited from 5G to 6G, ii) security threats of 6G architecture which result from architectural changes and new technologies used in 6G, and iii) security threats of 6G technologies, as stated below [Siriwardhana 2021]:

- Pre-6G Security Issues
    - Issues of 5G technologies inherited to 6G
    - Security and Privacy Issues of IoE
    - Security Issues of Local 6G Networks
- Security Threats of 6G Architecture
    - Tiny cells, Mesh Networks, and Multi-Connectivity
    - Sub-networks
    - RAN-Core Convergence
    - Zero-touch networks
    - Privacy Issues
- Security Threats of 6G Technologies
    - AI/DL
    - Blockchain
    - Quantum Computing
    - Visible Light Communication
    - Privacy Issues

***Analysis of the 5G and Beyond Threat landscape:***
It is too early to predict what security changes 6G will have exactly because it is not ready for testing yet. Still, from the theoretical analysis, the main challenges in 6G networks will be due to the use of modern technology. Several studies [Wright 2022, Duong 2022] have explained that recent technologies such as quantum computing, AI/DL and blockchain will be central components of 6G deployment. Thus, technology itself can be used to develop security solutions; for example, one exciting area is looking into Quantum machine learning (QML) [Kundu 2022]. QML can help expose the potential vulnerabilities of 6G, which can be used for developing dynamic attack models and adaptive countermeasures in protecting 6G systems.

## 4  AI/ML Cybersecurity in 5G and beyond networks

Artificial Intelligence (AI), and Machine Learning (ML) in particular, have in recent years evolved rapidly as disciplines. Consequently, ML is about to establish itself as the go-to methodology for all sorts of automation tasks for complex "systems" in a wide range of domains, including finance, media, entertainment, vehicle automation, natural language, and health. Cybersecurity and wireless systems like 5G are also in the process of entering this list. In this section, we give a brief overview of the ML components found in 5G and beyond the threats to these and provide some guidance on how to secure them. In particular, we discuss the use of ML for cybersecurity, with primary emphasis on intrusion detection.

### 4.1  ML components in 5G and beyond

As opposed to 4G systems, which rely on ML only sparsely, several parts of the 5G and beyond systems rely rather heavily on ML components. Kaur et al. [Kaur 2021] divide the 5G and beyond wireless network into three layers: 1) Physical layer (including sensing layer and network layer), 2) Middle layer (including the data link, network, transport, and session layer), and 3) Application layer. The authors argue that AI and ML will be embedded in all these layers and thereby play a key role in the system's performance. All the ML components are supposed to lead to more efficient, secure, and effective wireless communication.

In the physical layer, ML is supposed to support estimation, encoding and decoding in channels, as well as IP protocol configuration. For instance, Huang et al. [Huang 2018] propose a deep learning method for estimation of channel and so-called direction of arrival which is estimated through a combined offline/online training procedure. Another early example is [Joung 2016], applying kNN and SVM for optimizing antenna selection.

In the middle layer, ML will support authentication and logging services. ML may then be used to dynamically balance the server system. For instance, ML may be used for power saving by automatically setting servers and other system components to low-power states at non-peak hours, while rapidly turning off the low-power mode if traffic starts to pick up. Scheduling software updates to hours in which an ML model has predicted low traffic is another example.

In the application layer, ML will support file transfer, streaming services, etc. There, supervised learning can be applied to, e.g., intelligent caching and content prediction. For instance, Sadeghi et al. [Sadeghi 2017] use reinforcement learning in an intelligent content caching system that fetches popular content during non-peak hours. Moreover, unsupervised clustering techniques can be used to optimize network resources.

In most of the above ML applications, online learning techniques will be vital to adaptively adjust caching, channel tracking, load balancing, etc., to meet the requirements in the low latency applications typical in 5G and beyond.

## 4.2 Threats to ML components in 5G and beyond

Most of the literature on AI and ML related to 5G and beyond is concerned with the benefits, and opportunities the technology brings with it. While ML is expected to play a vital role in 5G and beyond, allowing a more efficient system, it also introduces a range of serious security vulnerabilities and issues. Suomalainen et al. [Suomalainen 2020] point out several weaknesses with a system that relies heavily on ML-based components. The threats to the ML components in 5G overlap with both threats to ML in general, and with general 5G and cybersecurity threats, and can be summarized as: DoS (Denial of Service), DoD (Denial of Detection, i.e. hinder ML from signaling attacks or failures], Unfair resource use (trick the ML into routing big resources to certain parts of the system) and sensitive data leakage (private or business-critical data revealed through e.g. learned models).

The very nature of ML has limitations that is important to be aware of. As ML models are trained on data, high quality data is required to get high quality ML models. For instance, to train an ML-based defense system against DoS, one needs realistic, relevant, and up-to-date data on DoS attacks. Further, as time goes on, user behavior may change, and the original data becomes outdated. This leads to an outdated ML model which requires retraining on new data and an ever-lasting maintenance burden. This is particularly critical for anomaly detection models. Moreover, since the models try to mimic the data, it is not seldom that minor changes to the input may lead to substantial changes in the output, causing unexpected situations and vulnerabilities. If the model is too tightly connected to sensitive training data, the data may be revealed based on the output of the model.

Intrusion detection systems (IDS) are often based on ML and are probably the most critical ML-based component in the 5G system. Due to its crucial role, and some of the threats to such ML components being very specific for IDS, we discuss this separately in section 4.5, along with other issues with ML in cybersecurity.

### 4.3 Data types

The data typically used to build and run ML-components in 5G and beyond, typically comes in the tabular data form, with fixed numeric (continuous or discrete) or categorical (unordered or ordered) features. In most cases there is a time component involved in the relevant data. Examples are live streams from network traffic, geographical positions of devices, and other system logs. See specific examples for IDS in Section 4.5.

### 4.4 Securing the ML algorithms

In view of the threats to the many ML components in 5G and beyond, a natural question to ask is how we can secure these algorithms?

First and foremost, it is critical that strict regimes are in place to ensure that both the data used for training the ML components, and the trained models themselves, cannot in any way be altered by unauthorized parties. With secure access to the data and models, data and model poisoning threats are minimized. It also secures privacy of the data.

Real, observational data are preferable over simulated data for training, as the latter typically only learn the rules with which the simulator was programmed and may also be out of date [Morocho-Cayamcela 2019]. Because of the requirement of real, recently observed data, proper data cleaning and filtering mechanisms must also be in place to ensure poisoned data does not enter the training data set. We also think it is problematic to rely on open data, as attackers can learn how models trained on these tend to work and exploit weaknesses in these data.

When training the models, one needs to pay attention to regularization, avoiding overfitted models. While this is problematic for performance, it can also be exploited by attackers to control the model output with minimal changes to the input. For models relying on sensitive data of any sort, one also needs to obey differential privacy, to avoid that the models implicitly contain sensitive data.

Once the models are trained, understanding how they behave and perform for different input data, their robustness, sensitivity, and extrapolation abilities, is important to understand the limitations of the model. Explainable AI (XAI) methods can help in this regard.

When ML components finally reach production, it will be crucial to have a well-functioning monitoring system in place, such that abnormalities in the use of the models can be detected and followed up. Since ML models are inherently non-transparent, simple rule-based fall-back methods should take over once abnormalities related to the use of ML models are detected. Furthermore, any services that reveal the output of the ML model are vulnerable to adversarial attacks. Thus, the model output should only be revealed where this is strictly required, otherwise, its access should be strictly controlled.

See also [ENISA 2021] for a broad taxonomic overview of ML algorithms from the security perspective.

## 4.5 ML for cybersecurity

Cybersecurity is the practice of protecting computer systems, networks, and data from unauthorized access, use, disclosure, disruption, modification, or destruction. Intrusion detection systems (IDSs) are a key component of cyber security, as they monitor network traffic and alert administrators to any potential security breaches or attacks. Machine learning (ML) is increasingly being used in IDSs to improve their accuracy. By using algorithms to analyze large amounts of data, machine learning can help an IDS identify patterns and anomalies that may indicate a security threat. This allows IDS to adapt and evolve to new threats and provide more comprehensive protection for computer systems and networks. In this section, which is heavily based on [Liu 2019], we review the use of ML in intrusion detection systems.

Machine learning has been applied to intrusion detection systems to help detect previously unknown attacks, known as zero-day attacks. Intrusion detection methods can be divided into two categories: *misuse detection* and *anomaly detection*. Misuse detection, also known as signature-based detection, uses a database of signatures of known attacks and raises an alarm when these are detected. This approach has a low false alarm rate, but it also has a high missed alarm rate because it can only detect known attacks that are in the database. Additionally, this database must be kept up to date, which can be costly. In contrast, anomaly detection uses machine learning to establish a model of normal behavior by learning from data. This approach can then raise an alarm when the observed behavior deviates significantly from the normal behavior. ML-based IDSs can detect zero-day attacks, i.e., previously unobserved attacks but have a higher false alarm rate and are less interpretable than misuse detection.

IDSs can be categorized as either *host-based* or *network-based*, depending on the type of data they monitor. Host-based IDS monitor logs from the host system's operating system and applications, while network-based IDS monitor network traffic in the form of *packets*, *flows*, or *sessions*. Packets are the basic units of network communication, while flows and sessions are aggregations of packets based on time windows and 5-tuple (client IP, client port, server IP, server port, protocol) grouping, respectively. The type of data used in an ML-based IDS may vary based on the types of attacks it is designed to defend against. For instance, flow-based methods may be more effective in detecting denial-of-service attacks that involve large amounts of traffic within a short time frame, while session-based methods may be more effective in detecting covert channel attacks that involve data leakage between two distinct IP addresses.

Machine learning-based intrusion detection systems (ML-IDSs) can be broadly classified into two categories: supervised and unsupervised methods. Supervised ML-IDSs use labeled data to train the system to classify incoming data as either an attack or regular activity. However, unsupervised ML-IDSs are trained to detect anomalies in raw, unlabeled data by learning typical patterns and issuing alerts when deviations from these patterns are observed. Although supervised methods tend to outperform unsupervised methods, they require labeled data, which can be expensive and difficult to obtain in real-world scenarios. While some open labeled datasets, such as *DARPA1998*, *KDD99*, *NSL-KDD* and *UNSW-NB15* [Moustafa 2015], are available, these datasets may not be relevant in practice due to their static nature and potential for obsolescence.

**Packet-based methods.** Packet-based methods classify individual packets as malicious or benign. Packet-parsing methods typically parse the protocol header fields and use these as features. A couple of examples are Mayhew et al. [Mayhew 2015] who group packets by protocol, cluster them by K-means++ and classify them using support vector machines (SVMs), and Hu et al. [Hu 2015] who apply an additional filter on alerts produced by Snort [Roesch 1999] using fuzzy C-means. Another approach is to model the packet payloads. Authors [Liu 2019b, Min 2018, Zeng 2019, Yu 2017] use deep learning algorithms to learn features from the payloads. Liu et al. [Liu 2019b] apply Convolutional Neural Networks (CNNs) and Recurrent Neural Networks (RNNs) to automatically learn features from packet payloads. Min et al. [Min 2018] combine statistical features and payload features learned using a CNN on the payloads to perform classification using Random Forest. Zeng et al. [Zeng 2019] use CNNs, RNNs and stacked autoencoders and select the best model based on validation accuracy. Yu et al. [Yu 2017] use unsupervised pre-training with a convolutional autoencoder to extract features and fine-tunes a classifier to predict attacks with labeled data.

**Flow-based methods.** Flow-based methods model a sequence of packets in a time window and can thus see packets in context of previous communication, something which is lost when processing only individual packets. Authors [Goeschel 2016, Kuttranont 2017, Peng 2018, Teng 2017, Ma 2016] engineer flow-based features such as average packet length, variance of packet length, ratio of TCP to UDP, the proportion of TCP flags etc. and use traditional ML methods such as support vector machines (SVMs), k-Nearest Neighbours (kNNs), Naïve bayes and decision trees to classify the flows as malicious or benign. Other authors [Potluri 2018, Zhang 2018] on the other hand, use deep learning to learn features directly from the flows. Potluri et al. [Potluri 2018] transform the flow into an 8x8 image and use a CNN to classify this image. Zhang et al. [Zhang 2018] learn features in an unsupervised fashion using a sparse autoencoder and use these features as input to an XGBoost classifier.

**Session-based methods.** Session-based methods model communication between two IP addresses at specific ports and a given protocol. Authors [Ahmim 2019, Alseiari 2015] use engineered features and classical ML methods. Ahmim et al. [Ahmim 2019] use a Random Forest on the features and the outputs of a rule-based model, while Alseiari et al. [Alseiari 2015] take an unsupervised approach where they cluster sessions before they manually label the found clusters. Features can also be learned directly from raw data using deep learning. Yuan et al. [Yuan 2017] perform DDoS detection using LSTMs, Radford et al. [Radford 2018] tokenize the communication in a session and learn word embeddings on these tokens which are used as input to a bi-LSTM, while Wang et al. [Wang 2017] used a character-level CNN on the bytes that make up the packets.

## 4.6 Practical challenges for ML-components in 5G and cyber security

As described above, the use of ML in 5G, particularly for cybersecurity, brings with it both possibilities and challenges. Morocho-Cayamcela et al. [Morocho-Cayamcela 2019] mention the scarcity of real data sets for training the ML models as one of the biggest challenges in the field. Liu and Lang [Liu 2019] mention three severe challenges for ML-based intrusion detection, and these should also be relevant for other ML-components:

*Lack of relevant labelled datasets*: If no relevant data is available and you use a supervised ML-based system, your only option will be to label data yourself. This can be an expensive process and would likely require labeling of new data that comes in at regular intervals to keep up with new attacks that emerge. Alternatively, you can rely on unsupervised ML.

*Low accuracy*: Most studies in literature are conducted using labeled datasets and performance might thus be limited in real environments where the incoming data only partially overlaps with the training data.

*Low efficiency*: Most studies focus on detection performance, whereas efficiency is not considered. However, for live deployment, efficiency is of utmost importance as a large amount of data must be processed in real time.

## 5   Concluding Remarks and Future Research Directions

In this study, we found out that 5G extends various networking opportunities because it offers more capacity, speed, and latency than its predecessors. The use of virtualized functions such as VNFs and VMs to achieve these opportunities introduces increased attack space. A recent report [ENISA 2022] pinpoints several security challenges that may occur due to VNF, NFV, VMs, and SDN. Therefore, advanced data analytics can be utilized for dealing with security threats (such as DDoS attacks) in 5G systems. New vulnerabilities in 5G and beyond systems due to NFV and SDN is also one of the main issues, AI/ML can be applied to propose potential solutions considering the dynamic and scalable properties of edge and IoT.

We also discovered that while the journey from 1G to 6G is about speed, the attack space has evolved from physical attacks to AI/ML and quantum attacks. The advantages of 5G and beyond are the unique combination of high-speed connectivity, low latency, and more effectiveness and efficiency than its predecessor technologies. All this development is, however, prone to security and privacy issues. Privacy violations may also occur during the training and testing phases of AI/ML attacks. We, therefore, agree with those who argue that efficient algorithms need to be provided by building adaptive algorithms on top of efficient model exchange by grouping, adaptive selection, and dynamic adjustment with minimal sacrifice on AI/ML model performance. Since most of the existing surveys on AI/ML 5G and beyond security mainly focus on the performance of AI/ML models and their accuracy, they often overlook the accountability and trustworthiness of the models' decisions. Therefore, future 6G security solutions must (i) allow transparent and comprehensible security decision-making processes making the systems accountable for automated actions, (ii) take a holistic approach addressing security, privacy, and trust which are closely related and will be critically important to the uptake and acceptance of 6G communication, and (iii) look in the lens of dynamism and adaptivity which will allow managing dynamic and evolving risks in 6G.

From our analysis, online learning techniques have the potential to play a prominent role in adapting, channel tracking, load balancing, etc., in 5G and beyond. Several challenges for ML-based intrusion detection need careful attention from the research perspective. IDSs must be able to adapt and evolve to offer dynamic security solutions in 5G and beyond.

Technology itself can be used to develop security solutions. For example, one exciting area is looking into Quantum machine learning (QML), which can help to predict potential vulnerabilities of 6G and develop dynamic adaptive protection of 6G systems.

**Acknowledgment**

This research has been supported by basic institute funding at Norsk Regnesentral, RCN grant number 342640, and the Research Council of Norway through the SFI Norwegian Centre for Cybersecurity in Critical Sectors (NORCICS), project no. 310105.